\documentclass{ws-ijmpe}

\begin{document}

\markboth{Johnson/Stetcu}{Shortcuts to nuclear structure}

%
\catchline{}{}{}{}{}
%

\title{Shortcuts to nuclear structure: lessons in Hartree-Fock, RPA,
and the no-core shell model}

\author{\footnotesize Calvin W. Johnson}

\address{Department of Physics, San Diego State University \\
San Diego, CA 92117, USA\\
cjohnson@sciences.sdsu.edu}

\author{Ionel Stetcu}

\address{Department of Physics, University of Arizona\\
Tucson, Arizona 85721, USA\\
stetcu@physics.arizona.edu}

\maketitle

\begin{history}
\received{(received date)}
\revised{(revised date)}
\end{history}

\begin{abstract}
While the no-core shell model is a state-of-the-art
microscopic approach to low-energy nuclear structure, its
intense computational requirements lead us to consider
time-honored approximations such as the Hartree-Fock (HF) approximation and
the random phase approximation (RPA).  We review
RPA and point out some common misunderstandings,
then apply HF+RPA to the no-core
shell model.  Here the main issue is appropriate treatment of
contamination by spurious center-of-mass motion.
\end{abstract}

\section{Introduction}

Don't believe everything you read.
Even recently written textbooks on introductory nuclear
physics can leave the reader with the impression that
our understanding of low-energy nuclear structure is
ill-defined. A typical example is found in Dunlap,\cite{dunlap2004}
who opens with the time-independent Schr\"odinger equation,
then states that for atomic and solid physics
``the difficulty...lies, not in our lack of understanding
of the fundamental properties of the electromagnetic
interaction, but in the complexity of the mathematics...
However, in nuclear physics the [potential] has not been
uniquely determined and a phenomenological approach is
usually adopted."  This statement is not really wrong,
but it is not really correct either\footnote{First of
all, because of the "complexity of the mathematics"
one almost always makes significant approximations
in atomic, molecular, and solid state calculations.
This footnote is too small to debate the justifications
for those approximations.}.

While it is fair to say that the nucleon-nucleon (NN)
potential has not been uniquely determined,
one can and should make a more careful statement: high-precision
data does uniquely determine the \textit{on-shell}
behavior of the the NN interaction. As discussed during this
symposium, it is the \textit{off-shell} behavior,
which is inextricably tied up with three-body on-shell
interactions, that is not determined. This important
distinction is often overlooked.

With the on-shell behavior of NN interactions firmly in hand,
the last 10+ years have seen a variety\cite{benchmark} of rigorous approaches to
microscopic, \textit{ab initio} calculations of the structure
of light nuclei. One of these is based on the well-known
interacting shell model, the ``no-core shell model'' of Bruce Barrett
and collaborators. It is an irony of history that
thirty years ago Barrett and Kirson's work\cite{BK70}) (on nonconvergence of particle-hole
corrections to effective interactions was one of
several bolides\cite{SW,VSW73} that caused a
``mass extinction'' of rigorous nuclear structure studies, by
showing serious flaws in the then-current
methodology.  Twenty years later, the no-core shell model was
proposed\cite{nocore} precisely to avoid those flaws.

We will not go deeply into the details\cite{nocoremethodology}
of no-core calculations,
but simply point out they are
heavily computational, requiring bases with
dimensions of millions, tens of millions, even hundreds of millions.
For reductionist, \textit{ab initio} calculations this is
the right thing to do. On the other hand, it is useful to seek
out  approximations that, first, allow one to get preliminary results
quickly and efficiently, and, second, illuminate the more detailed,
precise results.  Toward this end we turn to the Hartree-Fock
approximation, extended by the random-phase approximation.

In the next section we review some common understandings
about RPA, while in Section 3 we look at some common
\textit{mis}understandings about RPA. While some material
is a recitation of previous work, we present a new, nontrivial
calculation illustrating the ``collapse'' of RPA,
demonstrating the topic is more subtle than generally
appreciated. Finally, in Section 4 we present and discuss
preliminary results of applying
deformed HF+RPA to the no-core shell model.

\section{The Hartree-Fock and random phase approximations}

The Hartree-Fock approximation is based on the variational principle;
the trial wavefunction is a Slater determinant, an antisymmeterized product of
single-particle wavefunctions. (For a good
introduction we recommend the monograph of Ring and Schuck.\cite{ring})
The advantage of Hartree-Fock is that one can
interpret the many-body wavefunction in terms of single-particle
degrees of freedom. The disadvantage is that one loses correlations.

The random phase approximation is a generalization of Hartree-Fock that includes
small amplitude correlations.\cite{ring,Rowe70,BB94}
It can be derived by different approaches:
time-dependent Hartree-Fock, equations-of-motion,
and the quasi-boson harmonic approximation, which we favor.
There are two equivalent formulation of RPA: Green's function
and matrix. For the latter one solves
\begin{equation}
\left (
\begin{array}{cc}
\mathbf{A} & \mathbf{B} \\
-\mathbf{B}^* & - \mathbf{A}^*
\end{array}
\right  )
\left (
\begin{array}{c}
\vec{X}_\lambda \\
\vec{Y}_\lambda
\end{array}
\right )
=
\hbar \Omega_\lambda
\left (
\begin{array}{c}
\vec{X}_\lambda \\
\vec{Y}_\lambda
\end{array}
\right )
\label{RPAeqn}
\end{equation}
In simple terms, $\mathbf{A}$ is the sub-matrix of the Hamiltonian
$\hat{H}$ taken between one-particle, one-hole states, while
$\mathbf{B}$ is constructed from the matrix elements of $\hat{H}$
between the HF state and
two-particle, two-hole states. If one ignores $\mathbf{B}$, then one
is simply diagonalizing $\hat{H}$ in a truncated basis; this is
the Tamm-Dancoff approximation (TDA). TDA calculates only
excited states; RPA implicitly calculates corrections to
the ground state, but is not variational.

Hartree-Fock (HF) and the random-phase approximation (RPA) are old topics
in nuclear structure, and there is much lore about HF and RPA
in textbooks and monographs.
You might think there is nothing new to be
learned, at least not about the standard formulations.
Nonetheless, it is important to pay attention to several technical issues
if one wishes to apply HF+RPA to the no-core shell model:

 (1) \textit{Broken symmetries and their restoration.} Mean-field calculations
can break exact symmetries such as translational and rotational invariance.
By breaking an exact symmetry
one often gets a surprising improvement: for example, deformed solutions can be
lower in energy than spherical (rotationally invariant) solutions.

RPA and broken symmetries have a contentious relationship. It is
often stated that RPA ``restores'' broken symmetries.\cite{HB00,Lev04}
More careful and accurate statements can be found in
the literature: RPA``treats the inherent symmetries of
the problem consistently.''\cite{ring} What does this mean?
Eq.~(\ref{RPAeqn}) can be derived by a quadratic expansion
of the energy about HF state.
The vectors $\vec{X}_\lambda$ and $\vec{Y}_\lambda$ represent
particle-hole and hole-particle perturbations, respectively, on the
HF state; so for any perturbation corresponding to a generator of an
exact symmetry, for example a rotation, the energy
ought to be unchanged. In RPA the \textit{generators} of broken
symmetries are solutions of Eq.~(\ref{RPAeqn}) with $\Omega_\lambda = 0$.
By way of
contrast, TDA does not correctly identify broken symmetries as
zero-frequency
modes. It is critical to note that the zero-frequency mode can only
appear if the model space allows for exact restoration of
symmetries, a fact that will return to haunt us.

(2) \textit{``Collapse'' of RPA.} A salient issue is the so-called ``collapse'' of RPA.
RPA assumes small correlations, but in some calculations the
RPA corrections are unphysically large. This can be seen when the HF state
is near a transition from a symmetry-conserving state to a
symmetry-breaking state, for example,
from a spherical to a deformed state.
The transition can be driven by changing a parameter,
e.g. single-particle splitting.
The classic illustration  of collapse of RPA is the
Lipkin-Meshkov-Glick model.\cite{LipkinNPA,RowePR1968b}
Whenever one has symmetry breaking one worries
about unphysically large RPA corrections.

(3) \textit{Multi-shell calculations.}  In a large, multi-shell model
space, another symmetry that can be broken in HF is parity (i.e, by
mixing, for example, $s_{1/2}$ and $p_{1/2}$ single-particle
states). While most HF calculations enforce parity conservation,
some recent papers report HF with parity mixing.\cite{PNCMF} Because of the
possibility of ``collapse'' of RPA, however, it must be approached with
concern.A specific question is: which is more vulnerable to
collapse, deformation or parity mixing?

\subsection{Implementation of RPA in the interacting shell model: SHERPA}

We recently developed a code, SHERPA\cite{stetcuthesis} (SHell-model RPA), which
implements Hartree-Fock and RPA in
occupation space, solving the RPA matrix equations, Eq.~(\ref{RPAeqn}).
The only restriction
 is the Hartree-Fock wavefunction must be real; otherwise we allow for arbitrary
deformation and mixing as allowed by the model space.

Using the code SHERPA and comparing with exact results from
the Glasgow\cite{Whitehead77} and REDSTICK\cite{redstick} shell model codes,
we have written a series of papers carefully testing RPA in nontrivial
shell-model systems.\cite{stetcu2002,johnson2002,stetcu2003,stetcu2004}
The shell-model interactions used had dozens of
independent parameters, far more complicated than most previous tests
(the Lipkin model has only two independent parameters).

\section{Lessons from RPA}

In the course of applying SHERPA to the shell model, we found that many of
the common beliefs about RPA were not strictly true. Some  we discovered
through our calculations, others through a careful reading of the original
literature.

(1) \textit{Broken symmetries and ``restoration.''}
What the literature actually says is that
RPA yields an ``\textit{approximate} restoration of the symmetry''\cite{ring}
(italics added); the restoration is not exact, because the RPA wave
function is valid only in the vicinity of the HF state\cite{weneser}.
As a further investigation, we
computed RPA corrections to $J^2$ and other scalar operators.\cite{johnson2002}
For a deformed, even-even nucleus, the HF value of $J^2$ is nonzero.
The RPA corrections typically bring this value closer to zero, but not exactly;
and the RPA value of $J^2$ for the ground state can even be negative.
Such unphysical values arise because we have taken $J^2$ only to RPA order,
neglecting proper treatment of the Pauli principle, etc.

(2) \textit{Phase transitions and ``collapse'' of RPA}.  Long ago,
and mostly forgotten, Thouless
already correctly pointed out
two kinds of phase transitions.\cite{Th61}
In first-order transition, one has coexisting
solutions, each of which are locally stable HF, and
there is no collapse of RPA.  In a second-order transitions,
there is no coexistence, and collapse can occur. Thouless also
argues that a first-order transition will occur for even-parity modes,
e.g., quadrupole deformation, while
a second-order transition can occur for odd-parity modes.

We present  two illustations from the  shell model.  The first
example, shown in the right side of Fig.~1, is in the
$1s_{1/2}$-$0d_{3/2}$-$0d_{5/2}$ or $sd$ valence space with the
Wildenthal interaction,\cite{wildenthal} examining the transition
from deformed to spherical by lowering the $d_{5/2}$ single-particle
energy. We compute the RPA correlation energy\cite{stetcu2002} and
RPA corrections to scalar observables\cite{johnson2002}. 
(RPA is \textit{usually} an improvement over HF, 
but $Q^2$ in the spherical regime of $^{28}$Si is an exception.
For other scalars in $^{28}$Si RPA is an improvement for both spherical 
and deformed regimes\cite{johnson2002}.) 
RPA does not
collapse in this case: it is a second-order transition, because at
the transition point both the spherical and deformed HF solutions
are \textit{local} minima and thus stable.

\begin{figure}[th]
\vspace*{15pt}
\centerline{\psfig{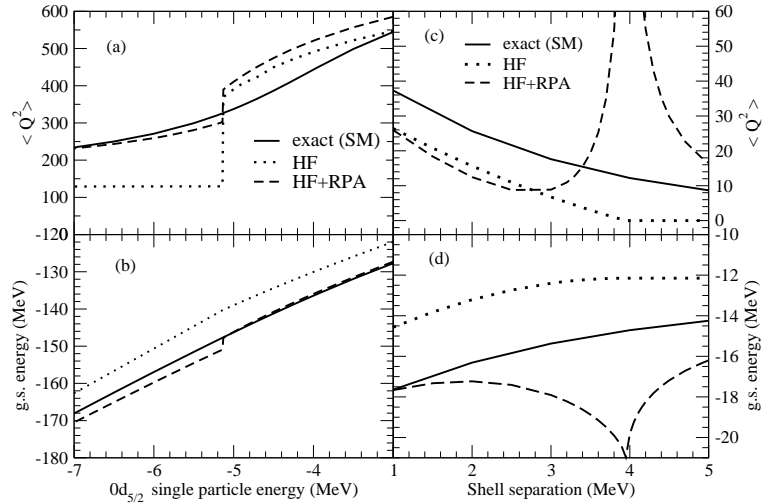}}\vspace*{8pt}
\caption{First- and second-order phase transitions in RPA. Left side
panels are $^{28}$Si in the $sd$-shell. As the $0d_{5/2}$
single-particle energy is lowered, the HF state transitions (first
order) from deformed to spherical; there is no collapse in either
$\langle Q^2 \rangle$ (a) or g.s. correlation energy (b). Right side
panels  are $^{16}$O in a $0p_{1/2}$-$0d_{5/2}$ space. As the
$0p_{1/2}$-$0d_{5/2}$ splitting increases, the HF state transitions
(second order) from mixed parity to pure parity. Both $\langle Q^2
\rangle$ (d) and g.s. correlation energy (d) show unphysical
contribution from, or ``collapse'' of, RPA. }
\end{figure}

In the second example, entirely new, we work
in the $0p_{1/2}$-$0d_{5/2}$ space with a combination
of interactions.\cite{wildenthal,ck,mk} Here the transition is between
HF states of good and mixed parity. (This turns out to be similar to
the Lipkin model, where the so-called
``deformed'' state is in fact a state of mixed parity.) We show
our results for $^{16}$O, but have similar results
for a variety of nuclides both in the $0p_{1/2}$-$0d_{5/2}$ space
and in a larger $0p_{1/2}$-$0p_{3/2}$-$1s_{1/2}$-$0d_{3/2}$-$0d_{5/2}$ space.
The RPA corrections grow unphysically large, a classic but
nontrival illustation of ``collapse.''


This demonstrates that, at least in an explicit $0\hbar\Omega$ space
a quadrupole shape transition is first order. It seems plausible
that Thouless' analysis will continue to hold in multi-shell spaces
and that the threat to multi-shell HF+RPA calculations will likely
come not from quadrupole deformations but from cross-shell,
parity-mixing (although we have not yet found explicit examples in
multi-$h\bar\Omega$ spaces).

\section{HF+RPA calculations for the no-core shell model}

We are not the first to apply HF to the no-core
shell model.  The pioneering calculations of Hasan, Vary, and
Navr\'atil\cite{HVN04} (HVN) looked at $^{4}$He and $^{16}$O in
spherical Hartree-Fock, with second-order corrections similar
to RPA.  They compared their mean-field calculations
to large-basis interacting shell-model (SM) calculations in a multi-shell
space: for $^4$He their full interacting shell-model calculations
included up to $10 \hbar \Omega$ excitations, while for $^{16}$O the
full calculations included up to $6 \hbar \Omega$ excitations.
Complete $N\hbar\Omega$ shell-model spaces, when used with
translationally invariant interactions, allows one to exactly
separate out spurious center-of-mass motion.

One of the main issues of applying Hartree-Fock to the no-core shell
model is the incongruency of model spaces. (If one
works in a $0\hbar\Omega$ space there is no incongruency.) In the interacting shell
model, one can limit the many-body basis to include all $N\hbar\Omega$
excitations. This is not possible for mean-field theory; instead one can
only define the single-particle space. For example, $^{16}$O in a
complete $4\hbar\Omega$ space includes, among others, 4p-4h excitations
from the $0p$ shell into the $1s0d$ as well as
1p-1h excitations from the $0p$
up to $2p1f0h$. But if the single-particle space for
the HF calculation includes the $2p1f0h$ shell, then the model
implicitly includes not only $4\hbar\Omega$ excitations but also
$6,8,10\ldots \hbar\Omega$ as well; but not \textit{all}
$10\hbar\Omega$ excitations, that is, the HF space is not a complete
$N\hbar\Omega$ space.

This has three consequences. First, it makes
comparison between the interacting shell model and
HF (+RPA) calculations problematic.
Second, there exist ``out-of-space'' two-body matrix elements
that appear in the HF space that
do not arise in the SM model space; what value
should one assign them? Third, because the HF+RPA  space is not a
complete $N\hbar\Omega$ space, spurious center-of-mass motion cannot
be separated out and will \textit{not} appear as a zero-frequency mode in
RPA.

HVN made several choices, all plausible but not inarguable. They
made the HF single-particle space smaller in extent than that for
the SM calculation (see their Figure 1).
Because of this they had few ``out-of-space'' two-body matrix
elements, and these were assigned, again plausibly but not
inarguably, the relative kinetic energy.
HVN obtained good values
for the ground state energies, although with large
second-order corrections.

Compared to HVN our calculations with SHERPA
have both an advantage and a disadvantage.
We allow for deformations (and odd numbers of nucleons,
although we did not exploit that here), so we can in principle
treat any light nucleus. This leads to a larger
computational burden, however, so that in our initial results described
here we could only tackle up to $2\hbar\Omega$ spaces.

We also made different choices for the HF+RPA space,
taking the same single-particle span as for the
SM calculation, and for the out-of-space two-body matrix
elements: we set them = 0.
We considered a number of $0s$- and $0p$-shell nuclei:
$^{4}$He, $^8$Be, $^{10}$B, $^{12,14}$C, $^{14}$N, and $^{16}$O,
all in a  $2\hbar\Omega$ space, using an effective interaction
derived from the bare Argonne $V8^\prime$
interaction.\cite{argonnev8}

\begin{figure}[th]
\vspace*{17pt} \centerline{\psfig{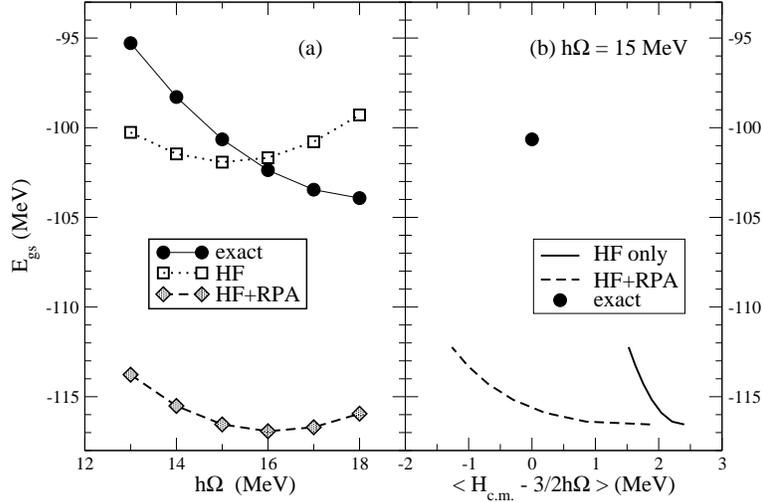}}
\vspace*{8pt} \caption{No-core calculation of $^{12}$C. Interacting
shell-model calculation (SM) in a $2\hbar\Omega$ model space; HF and
HF+RPA in a  $0s$-$0p$-$1s0d$-$1p0f$ single-particle space. Panel
(a) shows dependence on oscillator strength $\hbar \Omega$. In panel
(b) shows the results of constrained HF; see text for details. }
\end{figure}

We present our results in Fig.~2 for $^{12}$C; results for other
nuclides were similar. The ordinate axis is the value of $\hbar
\Omega$ for the harmonic oscillator basis, which sets the length
scale (in no-core methodology one typically scans $\hbar \Omega$ for
best convergence\cite{nocoremethodology}). The shell-model (SM)
calculations were performed in an exact $2 \hbar\Omega$ space. It
may seem surprising that even the HF values were lower than the
exact SM values, but this can happen because the HF calculation can
mix in spurious motion.



One signal of the mixing of higher-order excitations is that the
HF and RPA states will have spurious center-of-mass contamination,
signaled by  $\langle H_{c.m.} \rangle > \frac{3}{2}\hbar \Omega$.
Although we cannot exactly project out center-of-mass motion,
we tried constrained HF by adding
$\beta H_{c.m.}$ to our variational Hamiltonian. We then plotted the
ground state energy as a function of $\langle H_{c.m.} \rangle
-  \frac{3}{2}\hbar \Omega$;  extrapolating
the latter to value zero would then indicate the nonspurious component.
The results are shown in Fig.~2b, using a value of $\hbar\Omega = 15$ MeV,
and we computed $\langle H_{c.m.} \rangle $ in both HF and HF+RPA.

The results were not very good. In fact the HF+RPA value of
$\langle H_{c.m.} \rangle $ took on unphysical values $< \frac{3}{2}\hbar \Omega$.
This can happen for two reasons: approximations in RPA itself,
and the out-of-space matrix elements we chose to be zero.
In retrospect we strongly suspect the latter, especially when attempting to
apply constrained HF (where even for the added $H_{c.m.}$ we
kept the out-of-space matrix elements zero, with the unintended
consequence that one can no longer guarantee that
$\langle H_{c.m.} \rangle \geq 3/2 \hbar \Omega$; we did attempt
to remedy this but had problems with consistency we are still
attempting to resolve), and we plan to redo the calculations using
nonzero out-of-space matrix elements.

\section{Conclusion and summary}

We have discussed the Hartree-Fock and random phase approximations,
focussing on some hidden bits of lore regarding RPA, and
applied HF+RPA to the \textit{ab initio} no-core shell model.
In particular we have presented two new results:
(1) comparing, for the first time, in a non-trivial framework
(the interacting shell model) both first- and second-order transitions,
thus illuminating the so-called ``collapse'' of RPA; and (2)
applying HF+RPA to the no-core shell model with
arbitrary deformation and mixing of parity.  The latter is not
as successful as previous, spherical calculations, but the failure
is likely due to inadequate treatment of center-of-mass and
matrix elements in incongruenty model spaces.

\section*{Acknowledgements}

This work supported by Department of Energy grant DE-FG02-03ER41272
and by NFS grants PHY-0070858 and PHY-0244389.  The authors would like to thank James Vary
for enlightening discussions following presentation of this work.

\end{document}